   \documentclass[prl,aps,twocolumn,showpacs]{revtex4}
\usepackage[dvips]{graphicx}
\usepackage{dcolumn}%
\usepackage{amsmath}%
\usepackage{amsfonts}%
\usepackage{amssymb}
\providecommand{\U}[1]{\protect\rule{.1in}{.1in}}
\newcommand{\be}{\begin{equation}}
\newcommand{\ee}{\end{equation}}
\newcommand{\ba}{\begin{array}}
\newcommand{\ea}{\end{array}}
\newcommand{\nn}{ \nonumber}

\begin{document}

\title{The effect of Coulomb interactions on thermoelectric properties of quantum dots}

\author{Natalya A. Zimbovskaya}

\affiliation
{Department of Physics and Electronics, University of Puerto 
Rico-Humacao, CUH Station, Humacao, Puerto Rico 00791, USA; \\
 Institute for Functional Nanomaterials, University of Puerto Rico, San Juan, Puerto Ruco 00931, USA} 

\begin{abstract}
Thermoelectric effects in a quantum dot coupled to the source and
drain charge reservoirs are explored using a nonequilibrium Green's
functions formalism beyond the Hartree-Fock approximation. Thermal transport is analyzed within a linear response regime. A transition from Coulomb blockade regime to Kondo regime in thermoelectric transport through a single-level quantum dot is traced using unified approximations for the relevant Green's functions. 
   \end{abstract}


\date{\today}
\maketitle

Presently, thermoelectric properties of nanoscale systems have been attracting an increasing interest of the research community. In part, this interest arose because these materials  are expected to be useful in building up efficient energy conversion devices. Also, studies of energy (heat) transfer through  nanoscale systems can provide means for deeper insight into the nature and characteristics of general transport mechanisms, as was suggested in several works (see Refs. \cite{1,2,3,4,5}). Accordingly, in the last decade thermoelectric properties of diverse nanostructures such as quantum dots (QD), metal-molecule junctions, carbon nanotubes and other have been explored both experimentally \cite{6,7,8,9,10,11} and theoretically \cite{12,13,14,15,16,17,18,19,20,21,22,23}.

An important  characteristic feature of energy (heat) transfer through quantum dots and molecules is that under certain conditions, thermal conductance of such systems may be mostly determined by the  contribution from charge carriers whereas the phonon contribution remains comparatively small. For instance, electron contribution to thermal conductance may predominate at low temperatures \cite{20}. Also, it may happen that a significant mismatch occurs in the elastic properties of the QD/molecule and of their ambiance, thus implying a small overlap between phonon modes belonging to the former and latter subsystems and consequent predominance of the charge carriers contribution to the thermal conductance over the phonon part \cite{21}. Basing on the these and other conclusions and observations \cite{7,9}, phonon contributions are often disregarded in theoretical studies of thermal transport through QD and molecules  \cite{1,2,3,5,13,14,15,16,17,22,23,24}. In the present work we adopt the same approximation reducing thermal transport to its electronic component. 

It was shown that Coulomb interactions between charge carriers may strongly affect thermoelectric coefficients of QD/molecules leading to novel and distinct phenomena \cite{4,13,14,15,16,17,22,23}. In the last decade, these effects were (and still are) intensively studied.  In this work, we contribute to these studies by tracing the transition from the Coulomb blockade regime to Kondo regime in the characteristics of a single QD attached to the source and drain electron reservoirs.  As known, manifestations of Coulomb interactions between the charge carriers on the dot vary depending on relation between the charging energy $ U $ and coupling strength $ \tau $ characterizing the dot-leads contacts. The Coulomb blockade occurs when the dot is weakly coupled to the leads so that $ U $ significantly exceeds both $\tau $ and the thermal energy $k_BT \ (k_B$ being the Bolzmann's constant). Within the Coulomb blockade  regime, the equilibrium density of states of electrons (DOS) on a single-level QD displays two peaks whose separation equals $ U, $ as shown in the Fig. 1.

\begin{figure}[t] 
\begin{center}
\includegraphics[width=9cm,height=4.5cm]{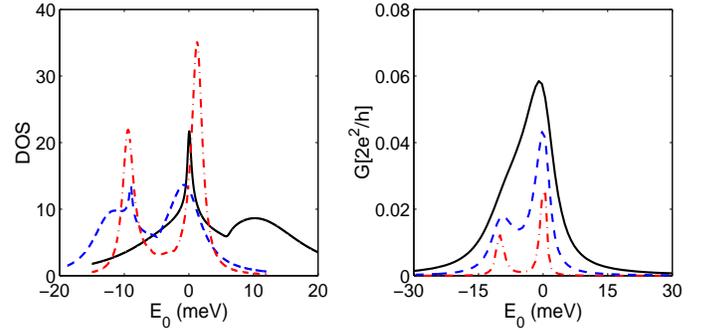}
\caption{(Color online) The equilibrium electron DOS on the quantum dot (left) and electrical conductance through the dot (right) versus the dot level energy $ E_0.$ The curves are plotted assuming  $ k_BT=0.26meV,\  U= 10meV,\ \tau =1.5meV $ (dash-dotted line),  2.5meV (dashed line) and 5meV (solid line), respectively. 
}
 \label{rateI}
\end{center}\end{figure}

These peaks give rise to maxima in the electric conductance occurring at the same values of $ E_0 $ which shown in the right panel of the figure. For a symmetrically coupled QD the peaks in the DOS as well as those in conductance have different heights. These differences are consistent with the shape of the corresponding current-voltage curves. These curves exhibit two steps, the first one being two times higher than the second. The ratio of heights has a clear physical sense. Obviously, one may put an electron of an arbitrary spin orientation to the empty level on the dot. However, the spin of a second electron added to the dot is already determined by the spin of the first one. So, there are two ways for adding/removing the first electron to/from the dot but only one way to do the same with the second electron. 

 As the coupling of the dot to the leads strengthens, the peaks in the DOS become lower and broader. When $ \tau $ growth further, the peaks disappear being replaced by a single sharp Kondo maximum responsible for the enhancement of electrical conductance of the dot at zero bias voltage.  This indicates  transition of the system from the Coulomb blockade transport regime to Kondo regime.
Both charge and energy transfer through a QD weakly coupled to the leads (Coulomb blockade regime) could be successfully analyzed employing quantum master equations or nonequilibrium Green's functions formalism (NEGF) within the Hartree-Fock approximation. These methods applied to a single-level QD symmetrically coupled to the leads give a correct description of current-voltage characteristics maintaining $\bf 2 : 1 $ ratio of heights of the successive steps \cite{25}.


It is well known that one cannot employ these relatively simple methods to study charge/energy transport through a QD/molecule strongly coupled to the leads. To catch the Kondo peak and analyze related transport phenomena one needs to employ  an advanced NEGF enabling to compute relevant Green's functions far beyond the Hartree-Fock approximation. A suitable for this purpose approach within the NEGF is based on the equations of motion method (EOM). Using this technique, one can express lower-order Green's functions is terms of higher-order Green's functions. As a result, one arrives at the infinite sequence of linear equations gradually involving Green's functions of higher orders. In outline, this procedure is well known and commonly employed. However, specific approximations used to truncate the set of EOM and to solve them vary which brings differences in the results for the Green's functions. Accordingly, the resulting response of the QD also vary.

This  may lead to inaccuracies in the obtained results arising from possible inaccuracies and errors in the employed approximations for the relevant Green's functions. These inaccuracies may not be easily disregarded for their existence makes questionable general results based on the advanced techniques within the NEGF formalism. 
For instance, Swirkowicz {\it et al} \cite{16} reported results of NEGF based calculations which give  two peaks of equal heights in the dependencies of both electrical and thermal conductances of the QD level position $E_0. $ This implies that current-voltage characteristics computed using approximations for the Green's functions employed in this work should reveal two steps of equal height, and correct heights ratio $\bf 2:1 $ is not observed.
  Similar difficulties in the analysis of charge transport through QD were discussed in an earlier work \cite{26}. There, it was suggested that acceptable approximations for the Green's functions intended for analysis of Kondo effect and relevant phenomena should be derived postulating an arbitrary relation between the parameters $U $ and $ \tau.$ Then the obtained approximations should be verified by tending to the limiting case of a weakly coupled system $(\tau\ll U). $   

In the present work, we adopt a similar approach to study the main thermoelectric characteristics of a QD represented by a single spin-degenerated energy level. For simplicity, we concentrate on electron contribution to the thermal transport omitting from consideration the phonon part. We write the relevant Hamiltonian as $ H = H_D + H_L + H_R + H_T $ where
the first term describes the dot  and is taken in the form
\be
H_D = \sum_\sigma E_\sigma d_\sigma^\dag d_\sigma + Ud_\sigma^\dag d_\sigma d_{-\sigma}^\dag d_{-\sigma}.        \label{1}
\ee
Here, $d_\sigma^\dag,d_\sigma $ are creation and annihilation operators for the electrons on the dot with a certain spin orientation, $ E_\sigma = E_0 $ is the energy of a single spin-degenerated dot level and $ U $ is the charging energy. The terms $ H_\beta\ (\beta = L,R) $ are corresponding to noninteracting electrons on the left/right electrode:
\be
H_\beta = \sum_{r\sigma} \epsilon_{r\beta\sigma} c_{r\beta\sigma}^\dag c_{r\beta\sigma} \label{2}
\ee
where $\epsilon_{r\beta\sigma} $ are single-electron energies on the lead $ \beta $ and $ c_{r\beta\sigma}^\dag,\ c_{r\beta\sigma} $ are creation and annihilation operators for electrons on the leads. The last term:
\be
H_T = \sum_{r\beta\sigma} \tau_{r\beta\sigma}^* c_{r\beta\sigma}^\dag d_\sigma + H.C.               \label{3}
\ee
describes tunneling effects between the dot and the electrodes. The factors $ \tau_{r\beta\sigma} $ are coupling parameters characterizing the coupling of the electron states on the dot to the leads. For a symmetrically  coupled system $ \tau_{rL\sigma} = \tau_{rR\sigma} \equiv \tau_{r\sigma}. $

Now, we employ the EOM method to compute the retarded Green's function for the dot. Disregarding spin-flip processes, we arrive at separate expressions for the Green's functions corresponding to different spin channels. These expressions can be presented in the form:
\be
G_\sigma^{rr}(E) = \frac{E - E_0 - \Sigma_{02}^\sigma - U(1 - \left<n_{-\sigma}\right>)}{(E - E_0 - \Sigma_{0\sigma})(E - E_0 - U - \Sigma_{02}^\sigma) + U\Sigma_{1\sigma}} .   \label{4}
\ee
Here, $\left< n_{-\sigma}\right> $ are one-particle occupation numbers on the dot:
\be
\big<n_\sigma\big> = \frac{1}{2\pi} \int dE \mbox{Im} \big(G_\sigma^<(E)\big) .  \label{5}
\ee
Self-energy corrections included into the expression (\ref{4}) have the form:
\begin{align}  
 \Sigma_{0\sigma} =\, & \sum_{r\beta} \frac{|\tau_{r\beta\sigma}|^2}{E - \epsilon_{e\beta\sigma} + i\eta} \equiv \Sigma_{0\sigma}^L + \Sigma_{0\sigma}^R, \label{6}
\\ 
\Sigma_{1\sigma} =  & \sum_{r\beta} |\tau_{r\beta,-\sigma}|^2 f_{r,-\sigma}^\beta 
\bigg\{\frac{1}{E -  \epsilon_{r\beta,-\sigma} + i\eta}
\nn\\ & +
 \frac{1}{E - 2E_0 - U + \epsilon_{r\beta,-\sigma} + i\eta}   \bigg\},  \label{7}
\\ 
\Sigma_{2\sigma} = &  \sum_{r\beta} |\tau_{r\beta,-\sigma}|^2 
\bigg\{\frac{1}{E -  \epsilon_{r\beta,-\sigma} + i\eta}
\nn\\ & +
\frac{1}{E - 2E_0 - U + \epsilon_{r\beta,-\sigma} + i\eta}  \bigg\}, \label{8}
\\  
\Sigma_{02}^\sigma =  &\, \Sigma_{0\sigma} + \Sigma_{2\sigma} \label{9}
\end{align}
where $ f_{r\sigma}^\beta $ is the Fermi distribution function for the energy $\epsilon_{r\beta\sigma}$ and chemical potential $ \mu_\beta $ and $ \eta $ is a positive infinitesimal parameter. The expression (\ref{4}) was first obtained by Meir et al \cite{27}. Later, the same expression was derived and employed in several works where transport properties of quantum dots and molecules were theoretically studied. 

The lesser Green's function $G_\sigma^<(E) $ is related to the retarded and advanced Green's functions ($ G_\sigma^{rr} (E) $ and  $ G_\sigma^{aa} (E) ,$ respectively) by Keldysh equation:
\be
G_\sigma^<(E) = G_\sigma^{rr}(E) \Sigma_\sigma^< (E) G_\sigma^{aa} (E).
\label{10}  \ee
In further calculations, we approximate $ \Sigma_\sigma^< (E) $ as follows:
\be
\Sigma_\sigma^< (E) = i \sum_\beta f_\sigma^\beta(E) \Gamma_\sigma^\beta (E) \label{11}
\ee
where $ \Gamma_\sigma^\beta(E) = - 2\mbox{Im} \big(\Sigma_{0\sigma}^\beta(E)\big) $ and $ f_\sigma^\beta (E) $ is the Fermi distribution function for the left/right electrode. This approximation was repeatedly employed in studies of thermal transport through a QD within the Coulomb blockade regime \cite{22,23}. Also, it was shown that expressions (\ref{4})--(\ref{11}) may be used to successfully trace the transition from Kondo regime to Coulomb blockade regime in the characteristics of charge transport through a single-level QD \cite{26}. The electric current through a symmetrically coupled system $\big[\Gamma_\sigma^L(E) = \Gamma_\sigma^R(E) \equiv \Gamma(E) \big]$ is given by the Landauer formula: 
\be
I = \frac{e}{\pi\hbar} \int T(E) \big[f^L(E) - f^R(E)\big] dE \label{12}
\ee 
where the electron transmission function has the form
\be
T(E) = \frac{i}{2} \Gamma(E) \big[G^{rr}(E) - G^{aa}(E) \big]. \label{13}
\ee

Thermal and electric properties of QD/molecule based devices are described by thermoelectric coefficients relating the charge current $ I $ and heat flow $ Q $ to the electrical bias $ \Delta V $ and temperature difference $ \Delta T $ applied across the system. In the linear temperature and bias regime, measurable thermoelectric coefficients are described by the expressions \cite{28}:
\begin{align}
G & = \frac{2e^2}{h}L_0,  \label{14}
\\
S & =- \frac{1}{eT}  \frac{L_1}{L_0}, \label{15}
\\
\kappa & = \frac{2}{hT} \left(L_2 - \frac{L_1^2}{L_0}\right),  \label{16}
\\
L_n & = - \int (E - \mu)^n T(E) \frac{\partial f(E)}{\partial E} dE.        \label{17}
\end{align}
Here, $ G $ and $ \kappa $ are electrical and thermal conductances, respectively, $ S $ is the thermopower (Seebeck coefficient) and $\mu = \mu_L = \mu_R $ is the chemical potential of the electrodes in the unbiased system.

As known, thermoelectric properties may be utilized in either charge-driven cooling devices or heat-driven current generators. The efficiency of both kinds of devices is determined by a dimensionless thermoelectric figure of merit $ ZT. $ Within the linear in $\Delta V $ and $ \Delta T $ approximation the figure of merit equals:
\be
ZT = \frac{S^2 GT}{\kappa} = \frac{L_1^2}{L_0L_2 - L_1^2}. \label{18}
\ee 

\begin{figure}[t] 
\begin{center}
\includegraphics[width=9cm,height=4.5cm]{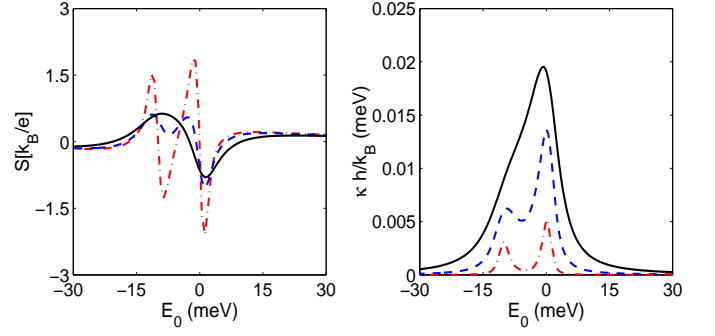}
\caption{(Color online) Thermopower (left) and thermal conductance (right) versus $ E_0. $ The curves are plotted assuming $  k_BT=0.26meV,\  U =10meV,\ \tau =1.5meV $ (dash-dotted line), 2.5meV (dashed line) and 5mev (solid line), respectively.  
}
 \label{rateI}
\end{center}\end{figure}

The electrical conductance, thermal conductance and thermopower are shown in the Figs. 1,2 as functions of the the dot level position $E_0$ for three different values of the coupling constant $ \tau. $ Calculations were carried out in the low temperature regime $ k_B T\ll U $ assuming $ k_B T = 0.26 meV, \ U = 10 meV. $ As discussed before, for a weakly coupled system $(\tau = 1.5 meV) $ the electrical conductance reveals two distinct peaks corresponding to Coulomb blockade peaks in the electron DOS. The peaks appear at $E_0 =0 $ and $ E_0 + U = 0 $ when either the original dot level $ E_0 $ or accompanying level $E_0 + U $ crosses the Fermi level in the electrodes $ \mu $ (assuming that $  \mu = 0). $ 
At stronger dot-lead coupling, $(\tau = 2.5meV) $ the peaks partially merge, and when $ \tau $ becomes comparable with $ U, $ a single peak emerges at $ E_0 = 0 $ which originates from the Kondo maximum in the electron density of states. Thermal conductance $ \kappa $  behaves similarly to the electrical conductance. For a weakly coupled system, two peaks are displayed at the dot level energies $ E_0 $ close to zero and $ - U ,$ respectively. These peaks become merged into a single Kondo maximum as the QD becomes sufficiently strongly coupled to the leads. 

The character of the thermopower dependence of the energy $ E_0 $ is more complicated. Within the Coulomb blockade regime, $ S $ rather sharply varies over the interval $ - U < E_0 < 0 $ changing sign three times. The specifics of this behavior may be explained as follows \cite{16}. When $ E_0 $ approaches the chemical potential of the leads from above, electrons start to tunnel from the electrode kept at higher temperature to the electrode of a lower temperature. Under the condition of vanishing current, the voltage drop is generated in the system. Due to the negative sign of electron charge, the thermopower measured in the units of $ k_B/e $ takes on negative values. As the decreasing dot level energy reaches the value corresponding to that of chemical potential, the electron flow becomes balanced by the holes flow, and the charge current disappears resulting in zero value of $ S. $ As $ E_0 $ further decreases, the net charge current and thermopower change the sign. Similar processes bring another resonance feature emerging near $ E_0 = -U. $ So, the thermopower changes its sign thrice: at $ E_0 = 0,\ E_0 = - U $ and in the middle of the Coulomb gap when $ E_0 = - \frac{1}{2}U. $
However, for strongly coupled QD, the dependence of thermopower of the dot energy is significantly simplified. Now, it is determined by a single Kondo maximum in the electron DOS instead of two peaks typical for the Coulomb blockade regime. Accordingly, $ S $ changes the sign only once. In general,  computed results for the electron and thermal conductance and thermopower are consistent with previously reported results computed separately for QD weakly and strongly coupled to the electrodes (see e.g. \cite{15,16,29} and references therein).  However, on the contrary with previously reported results, the heights of the peaks appearing in the thermal conductivity at $ E_0 = 0 $ and $ E_0 = -U $ as well as magnitudes of the corresponding features in the thermopower differ. This difference originates from the specific properties of the Green's functions given by Eqs. (\ref{4}),(\ref{10}),(\ref{11}). These Green's functions are proven to give correct $ \bf 2 : 1 $ ratio of heights of steps in the current-voltage curves for symmetrically coupled systems \cite{26}.

The figure of merit $ ZT $ is controlled by  already discussed thermoelectric coefficients $ G,\ \kappa $ and $ S. $ Its behavior is shown in the Fig. 3. $ ZT $ turns zero at the same values of $ E_0 $ which correspond to zero values of thermopower. Within the Coulomb blockade regime this happens three times whereas within the Kondo regime only once at $ E_0= 0. $ Accordingly, within the Coulomb blockade regime, $ ZT $ reveals four maxima, and these peaks may be relatively high. As shown in the figure, maximum values of $ ZT $ may be of the order of $ 1 $ thus confirming potential usefulness of QD/molecules as elements of efficient energy conversion devices. Within the Kondo regime, the figure of merit becomes zero only once, and its maxima are broader and lower than in the previous case. 

\begin{figure}[t] 
\begin{center}
\includegraphics[width=9cm,height=4.5cm]{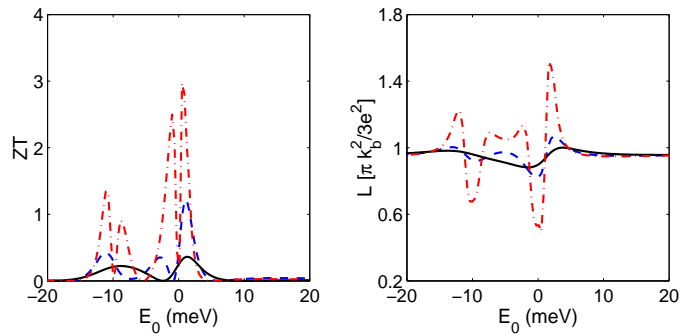}
\caption{(Color online)  Thermoelectric figure of merit ZT (left) and Lorenz ratio  (right) versus $ E_0. $ The curves are plotted assuming  $ k_BT=0.26meV,\  U =10meV,\ \tau =1.5meV$ (dash-dotted line), 2.5meV (dashed line) and 5mev (solid line), respectively.  
}
 \label{rateI}
\end{center}\end{figure}

As known, Coulomb interactions may lead to violation of Wiedemann-Franz law in mesoscopic systems. In QD/molecules, the Lorenz ratio $ L = \kappa/GT $ may significantly deviate from the standard value $ L^* = \frac{1}{3}(\pi k_B/e)^2 $, and these deviations may lead to an enhancement of the thermoelectric figure of merit \cite{14,19}. The dependence of Lorenz ratio of the dot energy $ E_0 $ for the considered system is shown in the right panel of Fig. 3 and noticeable violation of the Wiedemann-Franz law is displayed. We remark that comparatively low values of $ ZT $ and moderate manifestations of Wiedemann-Frantz law should not be misunderstood. Also there are no grounds to conclude that the Kondo regime is generally less favorable for providing high peak values of $ ZT $ than the Coulomb blockade regime. It is shown in earlier works \cite{14,18,22}, that $ ZT $ value is determined by the relation between three relevant energies, namely: $ k_BT,\ \Gamma, $ and $ U .$  Basing  on the approximations for the Green's functions given by Eqgs. (\ref{4})-(\ref{11}) and varying these energies one may show that $ZT $(as well as the Lorentz ratio) may reach   values  greater than unity within both Coulomb blockade and Kondo regimes provided that respective values of $ k_B T ,\ \Gamma $ and $ U $ reach optimal proportions. However,  the important problem of achieving the optimal efficiency in QD/molecule based thermoelectronic devices is beyond the  scope of the present work. It is discussed elsewhere \cite{5,14,15,16,22,23}.

Finally, in the present work we showed that the suggested computational scheme using a very simple approximation for the lesser Green's function $ G^<(E) $ brings appropriate results for the characteristics of thermal transport through QD/molecules within both Kondo and Coulomb blockade regimes. However, it is necessary to remark that these results are not exact. This follows from the specifics of the equations of motion (EOM) approach employed to compute the Green's functions. Within this approach, the Green's functions of lower order are expressed in terms of higher order Green's functions. Ultimately, one arrives at the infinite sequence of the equations successively involving Green's functions of higher orders. To derive  explicit expressions for the relevant Green's functions this system of EOM must be truncated. Also, higher-order functions still remaining in the system should be properly approximated. In principle, better approximations for the Green's functions used to analyze electrical and thermal transport through Kondo correlated systems these could be derived by applying a numerical renormalization group approach (NRG)  \cite{30}. 

Presently, NRG is the technique which brings  the most reliable results for transport characteristics of such systems, and it was used for this purpose in several works (see e.g. Ref. \cite{31}). Nevertheless, the EOM method has some favorable aspects which give grounds for numerous applications of the latter. First, the EOM based studies are less computationally costly than those based on NRG. Secondly the analytical expressions for the relevant Green's functions derived within the EOM method have simpler forms, which allows to easier analyze the effect of some important characteristics of the system (such as the relation between the energies $ \Gamma $ and $ U )$ on the transport properties. 
   The results presented in this work do not disagree with the corresponding NRG based results reported in Ref. \cite{31} as far as one can carry on the comparison. For instance, the dependence of low temperature $ (k_BT \ll \Gamma) $ thermopower of the gate voltage shifting the position of the dot energy level presented in Ref. \cite{31} has the form similar to that shown in the Fig. 2.
   
   The validity of the Green's functions adopted to study manifestations of Kondo effect in thermal transport  is verified by applying them to calculate characteristics of charge transport within the Coulomb blockade regime \cite{26}. Employing the approximation (\ref{11}) we imply that $ G^<(E) $ is not affected by Coulomb interactions on the QD as it is proved to be in the Hartree-Fock approximation. In principle, one may improve the approximation (\ref{11}) by taking into account corrections arising due to Coulomb interactions but these additional terms should disappear in the limit $ \tau/U \ll 1. $Therefore, we can reasonably assume that the present analysis brings correct results for QD not too strongly coupled to the leads so that $ \tau/U $ remains less than 1. We believe that the suggested computational scheme may be further developed to study characteristics of thermal transport through QD/molecules, beyond the linear approximation in bias voltage and temperature difference.
\vspace{2mm}

{\it Acknowledgments:} 
The author  thank  G. M. Zimbovsky for help with the manuscript. This work was partly supported  by  NSF-DMR-PREM 0934195 and UPRH-60FIPO490000.

  \end{document}